\documentclass[runningheads]{llncs}
\usepackage[T1]{fontenc}
\usepackage[utf8]{inputenc}
\usepackage{graphicx}
\usepackage{booktabs}
\usepackage{array}
\usepackage{siunitx}
\usepackage{amsmath}
\usepackage{hyperref}
\usepackage{color}

\usepackage{orcidlink}
\renewcommand{\orcidID}[1]{\orcidlink{#1}}

\begin{document}

\newcommand{\subhead}[1]{\par\vspace{3pt}\noindent\textbf{#1.}\ }
\title{AffectAI-Capture: A Reproducible Multimodal Protocol for Small-Group Meeting Research\thanks{Accepted at the AI4Action Workshop, 19th International Conference on PErvasive Technologies Related to Assistive Environments (PETRA 2026). Published by Springer in Communications in Computer and Information Science (CCIS).}}

\titlerunning{AffectAI-Capture for Small-Group Meetings}

\author{
Meisam Jamshidi Seikavandi\inst{1,2}\orcidID{0000-0002-1271-2481} \and
Alice Modica\inst{1,3} \and
Anna Obara\inst{1,4} \and
Fabricio Batista Narcizo\inst{1,2}\orcidID{0000-0003-1319-5153} \and
Tanya Ignatenko\inst{1} \and
Ted Vucurevich\inst{1} \and
Jesper B{\"u}nsow Boldt\inst{1} \and
Paolo Burelli\inst{2}\orcidID{0000-0003-2804-9028} \and
Andrew Burke Dittberner\inst{1}\orcidID{0000-0003-4985-7083}
}

\authorrunning{M. J. Seikavandi et al.}

\institute{
GN Advanced Science, GN Group, Ballerup, Denmark \and
IT University of Copenhagen, brAIn lab, Copenhagen, Denmark \and
Copenhagen Business School, Copenhagen, Denmark \and
Aalborg University, Denmark\\
\email{mejam@itu.dk}
}

\maketitle

\begin{abstract}
We present AffectAI-Capture, a protocol for collecting synchronized multimodal data in four-person meeting-like interactions, combining eye tracking, wearable physiology, close-talk and room audio, multi-view video, event logging, and structured self-report. Sessions use fixed task blocks grounded in established group-interaction paradigms, while acquisition and post-processing are organized around a single authoritative event timeline and standardized outputs. We describe the experimental rationale, synchronization philosophy, data organization, and practical trade-offs. Pilot-level validation of audio quality and video synchronization has been conducted using controlled bench tests; full protocol sessions with participants remain ongoing work. The contribution is a reproducible protocol architecture linking task design, instrumentation, timing provenance, and data packaging for affective, behavioral, and meeting-analytics research.

\end{abstract}

\section{Introduction}

Understanding group interaction in realistic meetings requires more than recording speech and video. Researchers increasingly seek to study how attention, arousal, effort, coordination, and subjective experience unfold jointly across participants and over time, demanding multimodal datasets that combine rich sensing, interpretable task structure, auditable synchronization, and reusable data organization.

Classic meeting corpora showed the importance of synchronized audio-visual capture and structured meeting scenarios for cross-session comparability~\cite{carletta2005_ami,janin2003_icsi}. More recent affective and physiological datasets demonstrated the value of eye-related and wearable measurements for studying internal-state variation, but are rarely built around true small-group meeting interactions~\cite{mirandacorrea2021_amigos}. This leaves a gap between meeting corpora optimized for collaborative structure and affective-computing datasets optimized for participant-level sensing.

AffectAI-Capture closes this gap by combining scenario-based small-group tasks with high-resolution multimodal sensing, explicit event-centered synchronization, and BIDS-oriented derived outputs. It builds on our prior work on gaze-based emotion perception, multimodal affect modelling, and the interplay of eye tracking, personality, and temporal dynamics in face-to-face settings~\cite{seikavandi2023gaze,j2024modeling,seikavandi2025modelling}, which highlighted the need for richer interaction-centered datasets. The protocol serves as a reusable foundation for affective computing, meeting analytics, and human-centered AI research.

\section{Design Goals}

The protocol is guided by four design goals.
\begin{enumerate}
  \item \textbf{Cross-session comparability without abandoning meeting realism.}
  Sessions are structured using fixed task blocks that reproduce distinct social-cognitive interaction regimes, while preserving a plausible overall meeting narrative.

  \item \textbf{Multimodal capture at both group and individual levels.}
  The protocol combines room-level behavioral signals with participant-specific channels such as gaze, pupillometry, skin conductance, and heart rate.

  \item \textbf{Auditable temporal provenance.}
  Synchronization is treated as a first-class design problem. The protocol does not assume perfect device synchronization, but instead records sufficient timing anchors to estimate, validate, and document alignment.

  \item \textbf{Reproducible and reusable outputs.}
  Vendor raw data are preserved, derived outputs are separated, and session outputs are structured in a BIDS-oriented manner to support downstream analysis and long-term reuse~\cite{gorgolewski2016_bids,pernet2019_eegbids,wilkinson2016_fair}.
\end{enumerate}

\section{Related Work}

Classic meeting corpora demonstrated the value of instrumented multi-party data, including structured scenarios and close-talk audio for transcription, overlap handling, and speaker-specific analysis~\cite{carletta2005_ami,janin2003_icsi}. However, these corpora focused on speech, video, and interaction annotation with limited access to participant-level measures such as gaze or physiology.

Multimodal affect datasets have shown the value of combining physiology, video, and self-report, but their setups often fell outside genuine small-group collaborative interaction~\cite{mirandacorrea2021_amigos}. K-EmoCon~\cite{park2020_kemocon} demonstrated the feasibility of combining wearable physiology and continuous annotation in naturalistic dyadic conversation, and AFFEC~\cite{sekiavandi2025advancing} introduced a multimodal dataset for face-to-face emotion communication integrating eye tracking, physiology, and video. Both bridge affective sensing and social interaction more closely than stimulus-response paradigms, but neither targets multi-party meeting settings with structured task variation. AffectAI-Capture is designed at the intersection of these traditions.

\section{Experimental Design}

\subhead{Session structure}
\noindent The protocol targets four-participant sessions organized into five blocks: T0 (onboarding/warm-up), T1 (hidden-profile decision), T2 (negotiation), T3 (idea generation and selection), and T4 (public-goods micro-game). The fixed order supports cross-session comparability while progressing through distinct demands: acclimatization, information pooling, strategic exchange, collaborative creativity, and cooperation under mixed incentives. Structured tasks are a scientific asset: they create repeated interaction contexts with known behavioral pressures, making multimodal measurements more interpretable~\cite{carletta2005_ami}.

\subhead{Task-specific rationale}
\noindent Each task block is included because it elicits a distinct and well-studied interaction regime.

\paragraph{T1: Hidden-profile decision.}
This task targets information pooling under asymmetrically distributed knowledge. It is suitable for studying whether groups disclose and integrate unique information, how attention and speaking patterns shift when new evidence is introduced, and how collective decisions are shaped by discussion structure~\cite{stasser1985_hiddenprofile,lu2012_hiddenprofile_meta}.

\paragraph{T2: Negotiation.}
Negotiation introduces strategic interaction, emotional signaling, concession dynamics, and interpersonal regulation. It is well suited for multimodal capture, as vocal, facial, physiological, and gaze-related signals may reflect tension, control, responsiveness, and the evolution of bargaining positions~\cite{vankleef2004_emotion_negotiation}.

\paragraph{T3: Idea generation and selection.}
This task separates divergent and convergent collaboration. It allows examination of how ideas are generated, shared, compared, and evaluated, as well as how groups converge on a final choice while weighing priorities such as originality, feasibility, or overall appeal~\cite{rietzschel2010selection}.

\paragraph{T4: Public-goods micro-game.}
This task introduces a controlled tension between private incentives and collective outcomes. It produces clear event boundaries for decision, reveal, and discussion, making it useful for event-locked multimodal analysis of cooperation, surprise, disappointment, and the negotiation of fairness and social norms~\cite{fehr2000_publicgoods_punishment,chaudhuri2011_publicgoods_survey}.

\subhead{Task timing}
\noindent Timed and moderator-controlled phases are intentionally mixed. Timed phases ensure comparability; controlled briefing phases preserve flexibility.

\begin{table}[t]
\centering
\small
\caption{Task and timing structure.}
\begin{tabular}{p{0.8cm}p{3.4cm}p{6.8cm}}
\toprule
Task & Objective & Timed phases \\
\midrule
T0 & onboarding & free talk: 300\,s \\
T1 & hidden-profile consensus & reading: 75\,s; discussion: 420\,s; selection: 60\,s \\
T2 & negotiation & negotiation: 480\,s; settlement form: untimed \\
T3 & ideation & generation: 180\,s; board \& discussion: 420\,s; selection: 60\,s \\
T4 & public-goods game & contribution: 60\,s; reveal: 60\,s; discussion: 180\,s \\
\bottomrule
\end{tabular}
\label{tab:tasks}
\end{table}

\section{Multimodal Instrumentation}

\subhead{Primary devices}
\noindent The stack includes 4 Tobii Pro Glasses~3 (eye tracking), 4 EmotiBit devices (physiology), 5--7 Jabra cameras, 5 DPA microphones (4 close-talk + 1 room/spare), optional Vicon motion capture, and 4 participant tablets with 1 shared display.

\subhead{Camera layout}
\noindent The video subsystem balances participant coverage, table-level interaction visibility, and synchronization robustness. The setup uses seven room cameras: four desk-adjacent cameras providing participant-focused lateral views and three overview cameras capturing the room more globally. Most cameras are configured for 1920$\times$1080 capture at 30\,fps, with the wide-angle room camera subject to device-dependent effective resolution limits in practice.

The layout is organized around a desk-centered spatial reference system. A permanent ChArUco board at the desk center defines the world origin, while desk-edge and participant-linked markers support calibration, participant localization, and later multimodal alignment. Four desk-adjacent cameras are physically mounted upside down to clear desk-edge obstructions and achieve downward-angled coverage of the participant's face and upper body from a compact mounting position; the resulting image is corrected in software to preserve consistent downstream orientation.

Capture settings prioritize temporal auditability. Cameras are recorded at a constant nominal frame rate, frame-level timing logs are preserved, and wallclock-based packet timestamps are preferred over device-local timestamps for cross-device consistency. Audio is captured separately rather than multiplexed into camera streams to reduce USB bandwidth contention and simplify synchronization diagnostics. The protocol prioritizes complementary viewpoints over a single canonical camera, since occlusion, head turning, and shared attention can make single-view capture insufficient in small-group meetings.

\begin{figure}[t]
    \centering
    \begin{minipage}[t]{0.47\textwidth}
        \centering
        \includegraphics[width=\textwidth]{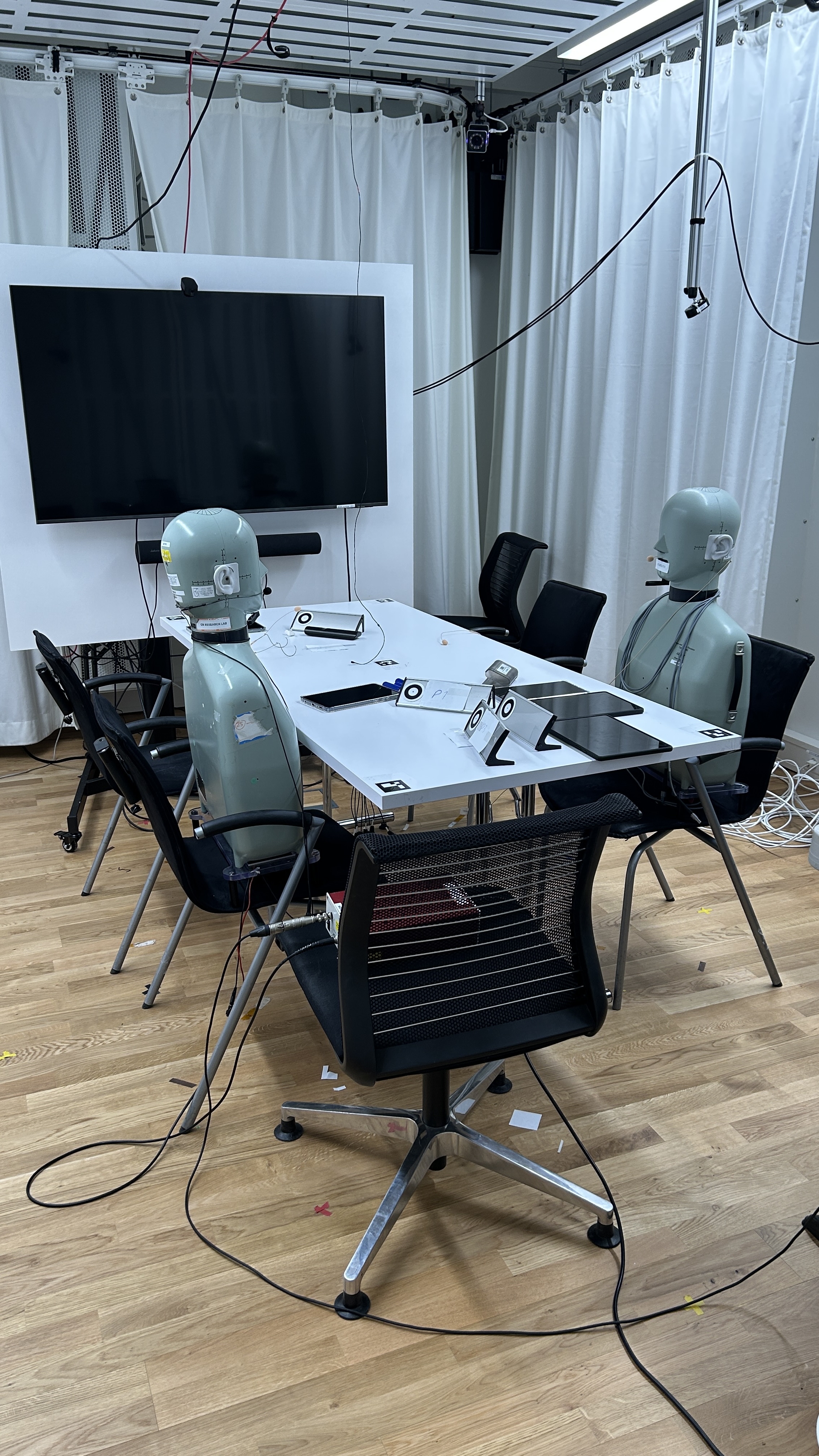}
        \caption{Audio validation using headset microphones on Head and Torso Simulators (HATS).}
        \label{fig:audio_check}
    \end{minipage}
    \hfill
    \begin{minipage}[t]{0.47\textwidth}
        \centering
        \includegraphics[width=\textwidth]{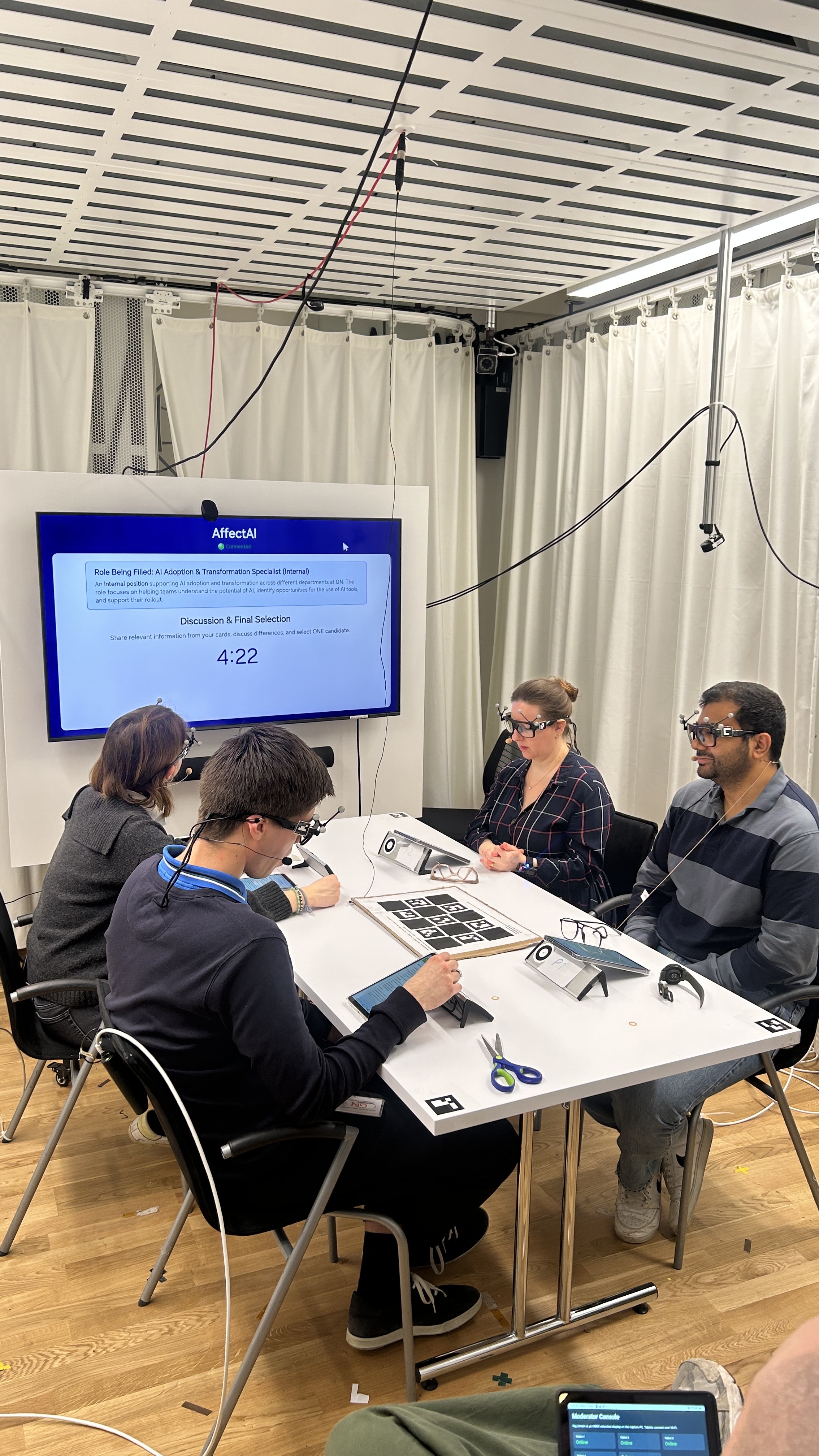}
        \caption{Pilot deployment of the AffectAI-Capture setup.}
        \label{fig:pilot_setup}
    \end{minipage}
\end{figure}

\subhead{Modality complementarity}
\noindent No single signal captures meeting dynamics adequately. \textbf{Audio} supports speech activity, overlap, diarization, and verbal content; \textbf{video} supports posture, gesture, and head orientation; \textbf{eye tracking} supports attentional allocation and interaction focus; \textbf{physiological signals} provide cautious proxies for arousal, effort, and autonomic activation; and \textbf{self-report} provides subjective grounding. These are treated as complementary indicators rather than direct readouts of psychological states, especially for pupil and physiological signals confounded by illumination, motion, and task context~\cite{mathot2018_pupillometry_review,posadaquintero2020_eda_review}. Recent work on multitask affective modelling from physiological signals further underscores the value of combining wearable channels~\cite{seikavandi2025mumtaffect}. Eye tracking is also a theoretically motivated channel for studying social-emotional processing~\cite{seikavandi2023gaze,j2024modeling,seikavandi2025modelling}.

\section{Self-Annotation Strategy}

Participants provide subjective labels through brief in-task valence--arousal--dominance prompts and post-task questionnaires capturing overall affect and experience with the task and interaction.

The rationale is twofold. First, self-report remains necessary when the goal includes subjective affect or perceived social experience, since behavioral and physiological signals alone cannot fully determine internal states. Second, in-task prompting allows labels to be linked to concrete phases and events rather than only retrospective summaries.

At the same time, prompt burden must be controlled to avoid turning self-report into the dominant task. Prompting is therefore event-contingent and phase-aware, with minimum spacing and condition-specific scheduling. The design follows the broader tradition of low-burden dimensional self-report, including valence--arousal--dominance style instruments and the Self-Assessment Manikin~\cite{bradley1994_sam}.

\section{Synchronization}

\subhead{Synchronization as a protocol concern}
\noindent In multimodal meeting studies, timing errors are not a minor implementation detail. They directly affect the interpretation of turn-taking, gaze-to-speaker relations, event-locked physiological responses, and self-report alignment. Accordingly, AffectAI-Capture treats synchronization as an explicit protocol concern rather than an invisible background service~\cite{anguera2012_diarization_review}.

\subhead{Authoritative event spine}
\noindent The protocol is centered on one authoritative session-level \texttt{events.tsv}. This file serves as the main temporal and semantic backbone of the session, recording task, phase, participant, stream, and event details in a machine-readable form. The goal is not to claim that all modalities are perfectly synchronized at acquisition time, but to ensure that alignment assumptions can be reconstructed and inspected later.

\subhead{Redundant timing anchors}
\noindent Synchronization relies on a hierarchy of anchors: LSL as the logical cross-device timebase, event logging at the protocol level, video frame logs and capture sidecars for camera timing, and modality-specific timing metadata when available. This redundancy is intentional: real acquisition systems fail in uneven ways, so the protocol records enough temporal evidence to validate and repair alignment post hoc rather than relying on a single fragile mechanism. This is consistent with LSL's role as a unifying synchronization layer for heterogeneous recordings, while acknowledging that device-specific latencies require explicit validation rather than assumptions of perfect synchrony~\cite{kothe2024_lsl}. Empirical characterization of inter-device offsets and clock drift across full sessions is planned as part of ongoing validation work and will be reported alongside the first complete dataset release.

\section{Data Organization}

AffectAI-Capture adopts a BIDS-oriented organization to make multimodal recordings easier to inspect, reuse, and process across heterogeneous streams. The main rationale is that raw data, metadata, events, and derived outputs should be clearly separated so that later analyses can reconstruct provenance rather than depend on undocumented preprocessing decisions~\cite{gorgolewski2016_bids,wilkinson2016_fair}.

At the session level, outputs follow a BIDS-like hierarchy with study metadata, participant/session folders, a session-level \texttt{events.tsv}, modality-specific subfolders, and a separate \texttt{sourcedata/} tree for vendor raw files. Raw data are preserved, while canonical and analysis-ready outputs are written separately. Derived products include modality-specific tables and media files, per-task run windows and event slices, normalized annotation outputs, and participant-signal mapping artifacts. This organization improves immediate usability while preserving the information needed for later re-analysis and audit.

As stable BIDS extensions for eye tracking and wearable physiology are not yet finalized, the current organization follows BIDS naming and hierarchy conventions while defining local modality-specific sidecars informed by draft extensions BEP020 (eye tracking) and BEP029 (virtual and physical motion data). These local conventions are documented in the dataset-level metadata so that future alignment with ratified extensions can be traced.

\section{Pilot Validation}

The validation work reported here consists of controlled bench tests for audio and video subsystems. No full protocol sessions with participants have yet been conducted; end-to-end validation under the complete protocol is part of ongoing work.

\subhead{Audio validation}
\noindent An audio quality test was conducted using four DPA d:fine CORE 4066 close-talk microphones connected to an RME Fireface~802 and recorded in REAPER. The goals were to verify gain staging, low noise floor, and minimal cross-talk behavior. The validation included playback of gray noise from a fixed source position, allowing measurement of signal-to-noise ratio across channels.

To simulate realistic participant conditions, two Head and Torso Simulators (HATS) were used (Fig.~\ref{fig:audio_check}). One HATS emitted either white noise or recorded speech, while the second wore a DPA microphone and was placed sequentially at different seating positions; the procedure was repeated for all microphones. Gain staging was set with conservative headroom, targeting approximately $-12$\,dBFS peaks during normal speech. Quantitative post-hoc analysis of SNR, noise floor variation, and recommended gain settings remains part of ongoing QC work.

\subhead{Video synchronization}
\noindent A multi-camera synchronization test used five Jabra cameras in the intended room configuration. The design principle was to record frame-level timing anchors and LSL-linked progress information during capture, then validate alignment offline using synchronized grid visualization across simultaneous views. This test was performed without participants present; the broader pilot deployment of the physical setup is illustrated in Fig.~\ref{fig:pilot_setup}.

\subhead{QC philosophy}
\noindent The broader QC strategy includes pre-flight checks for stream and device availability, fail-loud process supervision, session summaries and manifests, explicit reporting of missing streams and timing issues, and preservation of synchronization sidecars for later auditing. The aim is not to hide operational imperfections, but to document them in a way that preserves the scientific usability of the dataset~\cite{anguera2012_diarization_review}.

\section{Applicability to Sign Language and Assistive-Technology Research}

Although designed for spoken meetings, several architectural choices transfer to sign language and assistive-technology research. The multi-view camera rig and ChArUco reference frame are suited to capturing the signing space, since sign languages exploit the space in front of the body for referential and syntactic purposes~\cite{sandler2006_sign_linguistics}. The synchronization infrastructure and BIDS-oriented organization address challenges equally pressing in sign language translation (SLT) datasets, where temporal alignment between video, annotation, and auxiliary signals is critical~\cite{camgoz2018_slt}. The gaze instrumentation is relevant because eye gaze plays grammatical and turn-taking roles in signed communication, marking topic, focus, and turn transitions~\cite{baker1977_regulators}. These connections suggest AffectAI-Capture could serve as infrastructure for multimodal sign language and assistive interaction research.

\section{Limitations}

The protocol makes several deliberate trade-offs. Structured tasks improve comparability, but do not capture the full diversity of real-world meetings. Physiological and pupil-related signals are informative but indirect and must be interpreted with care. True temporal alignment across heterogeneous devices is never guaranteed by software alone, which is why redundant timing evidence is retained. Commodity camera stacks offer practical flexibility but can introduce platform- and bandwidth-specific instability, especially under USB-related hardware constraints. Some integrations also remain dependent on vendor tooling or device-specific constraints.

As noted above, the validation reported here is limited to controlled bench tests of individual subsystems; full end-to-end sessions with participants have not yet been conducted. Empirical characterization of synchronization accuracy, including measured inter-device offsets and drift, will accompany the first complete dataset release. These are not incidental weaknesses; they are part of the real design space of multimodal meeting science, and the contribution of the protocol is to make such constraints explicit and manageable.

\section{Conclusion}

AffectAI-Capture is a reproducible protocol for multimodal small-group meeting research, integrating structured task design, participant- and room-level sensing, synchronization-aware event modeling, and BIDS-oriented data organization. It provides a practical foundation for studying group affect, coordination, attention, and collaborative behavior, extending earlier work on gaze dynamics and emotion modelling~\cite{seikavandi2023gaze,j2024modeling,seikavandi2025modelling} into richer meeting settings. The architectural choices also transfer to sign language and assistive-technology research, where synchronized multi-view capture, gaze instrumentation, and provenance-aware data organization address open infrastructure needs.

\begin{credits}
\subsubsection{\ackname} This work was funded by GN Advanced Science, GN Group. Meisam Jamshidi Seikavandi received additional funding from Innovation Fund Denmark under the industrial postdoc programme.

\subsubsection{\discintname}
Meisam Jamshidi Seikavandi, Alice Modica, Anna Obara, Fabricio Batista Narcizo, Tanya Ignatenko, Ted Vucurevich, Jesper B{\"u}nsow Boldt, and Andrew Burke Dittberner are employed by or affiliated with GN Advanced Science, GN Group, which funded this research. Meisam Jamshidi Seikavandi received funding from Innovation Fund Denmark as an industrial postdoc. Paolo Burelli has no competing interests to declare.
\end{credits}

\bibliographystyle{splncs04}
\bibliography{references}
\end{document}